# Deep Learning and Health Informatics for Smart Monitoring and Diagnosis

Amin Gasmi

**Abstract**   The connection between the design and delivery of health care services using information technology is known as health informatics. It involves data usage, validation, and transfer of an integrated medical analysis using neural networks of multi-layer deep learning techniques to analyze complex data. For instance, Google incorporated "DeepMind" health mobile tool that integrates & leverage medical data needed to enhance professional healthcare delivery to patients. Moorfield Eye Hospital London introduced DeepMind Research Algorithms with dozens of retinal scans attributes while DeepMind UCL handled the identification of cancerous tissues using CT & MRI Scan tools. Atomise analyzed drugs and chemicals with Deep Learning Neural Networks to identify accurate pre-clinical prescriptions. Health informatics makes medical care intelligent, interactive, cost-effective, and accessible; especially with DL application tools for detecting the actual cause of diseases. The extensive use of neural network tools leads to the expansion of different medical disciplines which mitigates data complexity and enhances 3-4D overlap images using target point label data detectors that support data augmentation, un-semi-supervised learning, multi-modality and transfer learning architecture. Health science over the years focused on artificial intelligence tools for care delivery, chronic care management, prevention/wellness, clinical supports, and diagnosis. The outcome of their research leads to cardiac arrest diagnosis through Heart Signal Computer-Aided Diagnostic tool (CADX) and other multi-functional deep learning techniques that offer care, diagnosis & treatment. Health informatics provides monitored outcomes of human body organs through medical images that classify interstitial lung disease, detects image nodules for reconstruction & tumor segmentation. The emergent medical research applications gave rise to clinical-pathological human-level performing tools for handling Radiological, Ophthalmological, and Dental diagnosis. This research will evaluate methodologies, Deep learning architectures, approaches, bio-informatics, specified function requirements, monitoring tools, ANN (artificial neural network), data labeling & annotation algorithms that control data validation, modeling, and diagnosis of different diseases using smart monitoring health informatics applications.

**Introduction**

The fundamental use of deep learning in neural networks commenced as a result of experts study of complex neurons, layers, and its architectural paradigms. This study allowed to monitor data, its extended layer pipeline, and non-linear outputs generated from low-dimensional input space projection. Health informatics involve the generation of an automatic character set of human cells with expert intrusions. Medical imaging in health informatics can be elaborated to determine implicit internal organs like fibroids and polyps tissue irregularities. It can also be used to study morphological tumors (Fakoor et al.2013)[1]

Biologically, health informatics have anticipated translational features utilized for nucleotide DNA & RNA sequential protein strands. Convolutional neural nets (CNNs) as a deep learning approach in health informatics have architectural layers and filters for reducing, rectifying, and modifying poohing layers. The layers help to originate abstract features found in the visual cortex and receptive fields while other architectures like restructured Boltzmann machines, deep belief networks DBNs, stacked autoencoders, extended network and recurrent neural nets (RNNs); assist the advancement of graphical process units (GPUs) that impact the growth of deep learning applications.

Experts previously proposed pre-GPU and CNNs as parallel algebraic operations with matrixes needed for experiments in clinics. To integrate deep learning architectures in health informatics, its essential to label data and implement activation functions known as "transfer functions and weights". The transfer function must classify linear patterns to adjust the weights. McClelland et a. 1987 [3] proposed neural networks with hidden layers of perceptrons, stages, and epochs for new data input samples & weights with neurons adjustable based learning process, named "Delta Rule". The rule aids neural network training, exploitation, and backpropagation routines.

Rumelhart et al.1988 [4] observed the random values given to the network weights to determine it's iterative training processes and minimize the difference between network outputs and desired outcomes. Rumelhart et al.1988 [4] also furthered the study of iterative training using gradient descent techniques to reduce surface errors in the experiment. Deep learning accelerates supervised and un-supervised labeled data, whereby supervised labeled data train deep neural network to understand its weight, minimize errors, and predict the targeted value for classifying the unsupervised labeled data. Meanwhile, unsupervised deep learning data are utilized for clustering, reduction of dimension and featured extraction (Ngiam et al.2011)[2].

**Artificial Neural network and its variants**

Four deep learning architectures such as Auto-encoder, RBM, CNN, and RNN are mentioned to assist health informatics prediction and diagnosis. The autoencoder is referred to as feed-forwarder two-phase network that handles encoder and decoder tasks using input X and hidden H which represents non-linear equations stated below;

H - stand for non-linear activation functions that decodes maps hidden in the representation. Thus, the original hidden representation can also be calculated with

Z -If the model parameters optimize and minimize errors in the auto-encoder variants, the reconstructing error collection data = N, the sample square error optimization = $(X_i= f(x_i))2$

$X_i$ represents the ith sample of the unsupervised data, h stands for hidden representation and X for data sample (Bengio et al. 2007)[61].

The image below demonstrates the difference between physical based, conventional data-driven, and deep learning auto-encoder algorithm.

The conventional auto-encoder data-model requires handcrafting features for each individual trained module and cannot handle large datasets. Deep learning autoencoder methods provide end-to-end envisioned data structure without handcrafting features and train jointly large datasets;

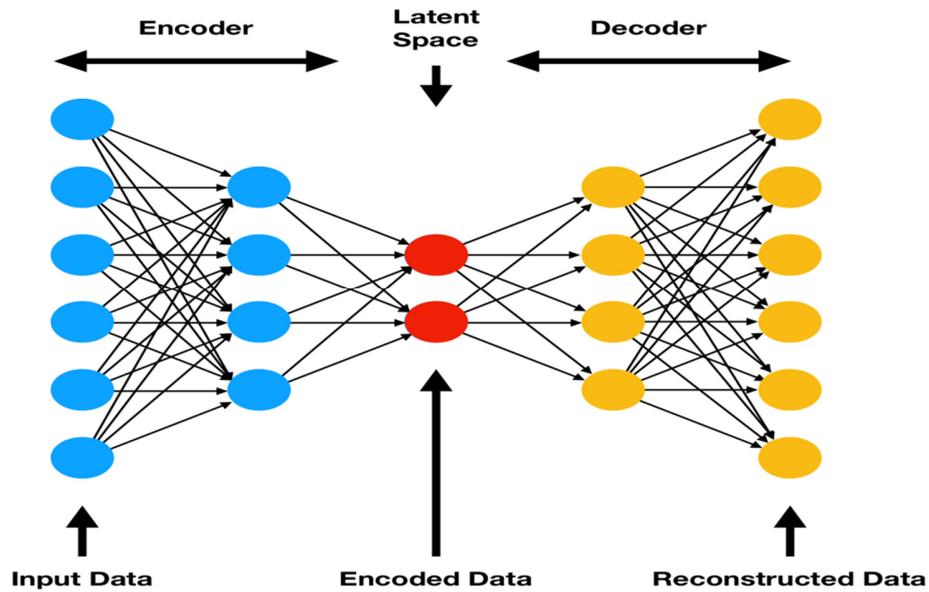

**Fig 1.** Autoencoder schematic illustration [78]

According to the diagram above, the learned transformation in the autoencoder must be sparse with constraints that implores the hidden unit with an optimizing function written below;

$(X_i = f(x_i))2$

M= hidden layer size

Ki=divergence hidden units and jth hidden neuron.

The conventional auto-encoder has an additional denoising network that corrects corrupt version of the data input, reconstruct, clean and train sample data X. it also has staking structure which puts together output lth layers as input (L+1)t-th layer to represent higher level and provide solutions to deep neural network model (Vincent et al.2008)[62].

**RBM variants**

Restricted Boltzmann machine has two-layers (NN) bipartite graph consisting of two visible groups known as units V and hidden unit h with an asymmetric link between both, but doesn't connect their nodes.

RBM model parameters = (w,b,a) energy functions.

Where = Wij –Vi- hj.

Wij -connecting weight between units

Vi- total number 1 and hidden unit

Hi -the total number of Ji bi and ai, which shows the joint RBM distribution over the unit based energy function equated as p(v, hj, Z -partitioned function/normalization factors).

The conditional probabilities of the hidden & visible units h and V will be

P (hj =I/Vj )

P (Vi=I/Vj = logistic function

The w-learning approach is achieved using a contrastive divergence tool (CD).

**Deep belief network variants**

DBN is made up of stack multiple RBM with output ith layer (hidden unit and input (L+1)-th visible layer. DBN has a common similarity with SDA due to its large layer unsupervised pattern in handling pre-trained data parameters.

Deep Boltzmann machine learning approach contains hidden units grouped into layers of single connectivity constraints with its full connection found between subsequent and non-neighboring layers.

## Convolutional neural network and its variants

CNN was utilized to perform image spatial processing via weights & pooling properties. CNN serves the purpose of authenticating natural language & speech recognition. It helps to learn about the alternating abstract features, stack convolutions & pool operations. The two dimensional CNN can be compared to a one-dimensional model using input sequence data $X=(X_i.....................X_t)$ where t represents lengths of sequence, Vi-d respectively.

The convolutional dot production can be filtered with vector U

Where $C_i$ =

Where XT stands for matrix x, b and the output $C_i$ is seen as the activated filter U that correspond with $X_i$; I + m-1. The slide filtering window features a map vector of $C_j$ =($C_1$, $C_2$, ........................(l-M+1)

The J represents the index J-th filter that corresponds with multi-windows ($X_i$; m, $X_2$: m+1................$X_{1-m}$ + 1:1).

CNN has a max-pooling layer that is capable of reducing the length of the featured map and minimized the numbers of modeled parameters. It has a hypermeter pooling layer denotation of MAX operation with consecutive value S and feature map $C_j$.

The compressed feature h= (h, $h_2$, --------- +1)

$H_j$= max ($c_{(j-1)}s$, C (J-1) S+1, ................C, S-1).

To predict data possibilities, the alternating CNN max-pooling layers can fully be connected to the softmax layer.

## Recurrent neural network and its variants

Schmidhuber (2015)[63] highlights the arbitrary length sequence of pattern input, which builds the connection between direct cycle and multi-layer perceptron. RNN trains back-propagated supervised data with subsequent input and targeted datasets (Jaeger, 2002)[64]. It's functional transition step T shows time information (Xt) moves from prior hidden output ht-1 to update the current hidden output ht = H (t,ht-1).

H - non-linear & differential transforming function.

Ht- learned representation showing input data and length T. Also, the conventional multi-layer perceptron mapped the obtainable ht representation to make the prediction successfully.

The simple function "Vanilla RNN", can be equated as-ht

W and H that represent transformation matrixes while b =bias vector.

Vanilla RNN suffers vanishing gradient problem due to back-propagation, but can easily be restored with LSTM and gate recurrent neural networks (GRU) to prevent errors and explosion. (Chung et al.2014)[65].

The advanced version of LSTM & GRUs has a multi-layer recurrent bi-directional model capable of offering structural flexibility.

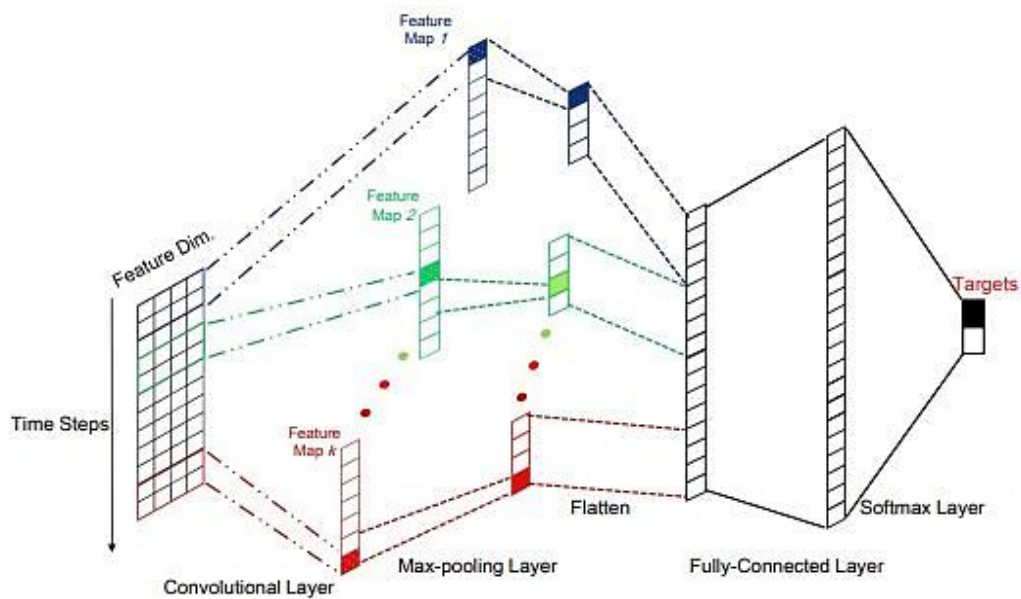

**Fig 2.** One-layer CNN, pooling layers, fully connected one Softmax Layer [79]

**Health monitoring applications and wearables**

Rose et al. 2010 [5] implemented hierarchical clustering methods to detect mammographic image data. The image below shows health monitoring applications necessary for capturing arrays of pervasive sensors worn or implanted in the body to capture ambient inertia motion, ECG patches, smart watches, EEG, and prosthetics (Johnson et al.2016) [6].

Pervasive sensors are wearables implanted as ambient sensors that monitor human health and accurately estimate food intake, energy expenditure, tackles obesity, chronic diseases, and care for patients with disabilities. Patients undergoing rehabilitation and critical care situation are often implanted with assisting devices to check vital signs (Pouladzadeh et al. 2016)[7].

During epidemics, the escalation of health-related issues like obesity, and cardiovascular diseases are controlled with energy expenditure/activity recognition tool, since it controls the amount of diet by monitoring calorie intake. CNN can alternatively be used to recognize and monitor accurately food intake by adopting cloud computing, size-calibrating, and distance estimation tools.

Pouladzadeh et al. 2016 [7] combined DL method with invariant hierarchical representation of video using two-layer 3D convolution & max-pooling large inputs to recognize human daily activities. Yalcin, 2016[9] used RGB D-video sequence to classify human activities and mount surveillance on elderly and child care clinics. CNN has furthered the detection of baby's fall & crawls while alerting caregivers by raising an alarm. The RBMs work together on smartphones & watches (Assisting devices), with audio & tactile feedback application, specifically used to detect visual impairment. Some assistive device consists of CNN DL algorithms that recognize hand gestures, sign languages, sterile surroundings, and permits touch-less human-computer-interaction (HRI). Huang et al.2015[8] introduced a deep neural network (DNN) for sign language recognition using real-sense data that coordinates finger joints inputs without handcraft features. DL machine ensures that health monitoring is achieved with discrete targeted values applied through Softmax later (prognosis and linear regression layer), that limits human labor and expert's skills.

Sun et al. (2016)[41] used a one-layer auto-encoder neural network to classify health-machine motor fault recognition and repair while Lu et al. (2017)[42] diagnose rotary health machine faults with components of stacked denoising AE in the three hidden notable layers. Most devices for vital signs like ECG, BCG (ballistocardiogram) are built with DL applications. Technology advancement brought about wearable photos or videos attached to outfits, and embedded sensors placed on chairs, car seats, and mattresses (e.g. Fitbit wrist band tracker controlled with mobile apps and add-ons).

Experts studied the use of a genetic triaxle algorithm known as accelerometer bracelets to trace walking patterns (e.g. falls and seizure inpatients). Research declared that prolonged sedentary lifestyle can cause adverse health outcomes which made clinicians adopt the use of wearable devices for monitoring patients' health and advice physical activity when necessary. Choo et al. 2017[76] used wearable devices to trace language patterns of mother-child connection and childhood psychological development. Choi et al. (2017)[75] monitored stress patterns using mentally equipped sensors powered by Ml/DL to understand stress in children and adults by following their heartbeat, blood pressure, and temperature.

According to our research, the electrodermal activity tool (EDA) known as the "emotion Board" has helped to measure skin reaction to stress. However, SVM and LDA help to classify stress and show up to 82.8% result.

Chen et al. (2018) monitored heat stroke using a fuzzy logic technique to display an inferential signal on smart devices. The design of wireless communication profoundly changed patient's management through point-to-care diagnostic devices like EMS (emergency medical service) used in emergency rooms & ICU. Some ICU systems revealed beyond vital signs to the extent of detecting patients' posture/position, toxic gases, and heat flux. Winokur et al. 2013[28] found wearable devices valuable for monitoring heart rate, recording ECG, 3D representation of Sternal Seismocardiogram (SCG).

Da He et al. 2012 [27] introduced ear device and wearable cardio-meter defibrillator WCD) to prevent arrhythmic sudden death of patients. Fryar et al. 2017(12) detect hypertension with physiological signals from cerebral blood flow-meter (CBF) that estimates hemodynamic cephalic symptoms in patients.

**Deep learning architectures, descriptions, and key points**

Deep learning networks are frameworks for classifying or regressing data which have either hidden or visible output layers. Some have more than two hidden layers that allow the expression of complex or non-linear hypothesis. Deep neural networks have successfully been used in bio-informatics but lack training authentication due to its back-propagated layers and slow learning process.

Hinton and Salakhutdinov, 2006 [10] discussed deep auto-encoder designed to extract features and later dimensional reduction with its input, hidden, and output layers. Deep auto-encoder consists of similar input & output nodes, which help to recreate input vectors and handles unsupervised learning techniques.

The deep auto-encoder is vital for labeling data, and robust representation (Sparse-AutEU). However, it does need pre-training to undergo the full training process. The deep belief network RBM composition has each sub-network hidden layers & visible layers with undirected connections. The two layers permit unsupervised & supervised training networks to initialize network commands, inference tracing for handling sampling process, and training procedures. (Hinton et al. 2016)[10].

Salakhutdinov and Hinton, 2009 [11] states the difference between deep belief network and deep Boltzmann machine network. According to his laid emphasis, Boltzmann has conditional independent layers that are undirected but uses stochastic algorithms to maximize lower bounds, incorporate robust inference and ambiguous inputs. Boltzman DL can also handle complex time inference that is higher than DBN; while optimizing large dataset parameters.

The continual progression in neural networks led to the proposition of recurrent neural networks with the capacity to analyze huge data streams, memorize sequential events, model time dependencies, and process natural languages. The recurrent neural network faces a challenge of vanishing exploding gradients. (Williams and Zipser, 1989)[13]. The convolutional neural network is communicably used since its quite compatible with 2D data and transforms filtered input to 3D output for neuron activations (LeCun et al. 1998)[14]. CNN supports Neuro-biological modeling, which performs visual cortex using flow neuron connections and many varied applications like Google Net & Clarifa. The main challenge in CNN is the hierarchical visual feature used to input large labeled datasets. The parallel GPU acceleration offers hardware capacity need to compute DNN on clouds and multi-core processors. The recurrent neural network comes with a hidden capacity to analyze several data, which made Bengio et al. (1994)[61] discuss RNN variation called Long-Short term memory unit (LSTM). LSTM solves gradient vanishing problems using long-input sequences.

LSTM can be used to exploit stored information, write and read information without errors during training. It's compatible with RNN and shares the same weight whilst aiding natural language processing like modeling, speech recognition, and image description.

Ackley et al. 1986 [16] emphasized on the variant Boltzmann machine (RBM) type of stochastic neural networks with Gaussian learning procedures called GIBBS sampling. GIBBS sampling adjusts weights, minimize errors, and model the variable relationship probability. Wang et al . 2016 [17] reviewed the graphical probability model with stochastic units and characterize the conditional independence between variables and directed acyclic graph.

Carreira and Hinton, 2005 [18]  mentioned the contrastive divergence algorithm (CD) used in conjunction with RBM to handle unsupervised learning algorithms. It has positive and negative phases, whereby the positive phase encourages network configuration and the negative phase recreates current network configurations. CNN has regular correlated local data with multi-dimensional input that can be significant in back-propagation, adjusting of number parameters that support the neuro-biological visual cortex model. The visual cortex in CNN has receptive local field maps that move granularity anterior image inputs to a convolved sub-sampled output through small filters (Hubel and Wiesel, 1962)[19].

**DNN learning architectures**

Deep neural network architecture known as input-output deep architecture (IODA) can resolve different image labeling issues by assigning labels to each image pixel. DNN services both hyperspectral images, whereas spectral & spatial features come together to form hierarchical models that characterize human body issues. Kondo et al. 2014[20] employed a group method of data handling (GMDH) with a hybrid multi-layer neural process to authenticate polynomial activated functions. The essence of GMDH is to recognize liver and spleen data while performing principal component regression analysis. The same technique can be used to identify Cancer Mcyarduim, right, and left kidney issues.

**Application of deep neural network to translational bioinformatics**

Table 1 demonstrates software explored with CUBA/NVIDIA to aid GPU acceleration which Wolfram Mathetica and Nervana (2020)[74] used to provide cloud training process systems in combination with neuromorphic electronic system hardware.

Most computational neuroscience simulations are conducted with neurons & synapses chips, integrated into hardware like (IBM) true north, (Spinnaker), and Curie (intel). The main purpose of bioinformatics as a discipline is to explore, investigate, and understand the biological molecular level and its processes. Past human genome project (HGP) researched raw data to develop new hypotheses of genes, and environmental factors related to the creation of human genetic proteins. To diagnose diseases using biotechnology, the first human genome motivating principle is "P4" (personalized, preventive, participatory, and predictive medicine) (Hood et al.2011)[22].

The predictive health informatics hold attributes for encoding DNA of the living beings while analyzes the alleles, environmental factors leading to diseases like cancer, and design targeted therapeutic procedures as a remedy. (Leung et al.2016)[23].

The concept of pharmacogenomics is focused on evaluating varieties of drugs & its response to gene-related treatment for aliments especially personalized diagnosis with fewer side effects. Epigenomics investigates interactive proteins and its higher-level processes, and response while transcriptome (mRNA), Proteome and Metabolome modify gene's response to its environments.

Genetic variants are uniquely designed with slicing codes that predict the differences in human tissues, especially how it changes as a result of genetic variation. The alternate of slicing code helps to technically generate gene prediction of slicing patterns meant to comprehend gene phenotypes and its drug effects on autism, spinal muscular atrophy, and hereditary cancer. (Leung et al.2016)[23].

A quantitative activity structure relationship was meant to predict protein-protein coordinations which are usually structured with molecular information, compound interactive protein; for predicting proteins used for drug discovery. Compound interactive protein virtual analysis influenced the discovery of new compounds, toxic substances, and the interpretation of drugs related to targeted cells.

In health informatics, deep learning models are utilized to enhance DNA methylation for providing visible outlooks of human chromosomes. It can be used to identify unstable chromosomes, error translation, cell transcription, differentiation, and cancer progression. (Angermueller et al.2016)[24].

Pastur-Romay et al. 2016 [25] named Chembl database in pharmacogenomics that detects millions of compounds descriptions used to develop & target drug evolutions, since the mentioned database encrypts molecular fingerprints, and understand traditional Ml approaches. Chembl database also helps to reduce data complexity in molecular RNA by binding predictive proteins together using RNA structural tertiary profiled outcome (Zhang et al. 2016)[49].

Fakoor et al. (2013)[1] utilized the autoencoder model to explore genetic data from diverse cancer patients to identify similar microarrays in the datasets. Ibrahim et al.2014 [26] detailed the effect of active learning methods using DBN to feature MicroRNA for classifying the performance of different cancer diseases like hepatocellular carcinoma. Deep learning approaches were adopted by Khadem et al. (2015)[28] to beat breast cancer disease through an attribute & noise combination of BDN & Bayesian network that helps the extraction of micro-array data. Experts considered deep learning more effective than SVM in detecting slicing code of different genetic variants, which according to Angermueller et al. (2016)[24] DNN predicts DNA methylation from an incomplete sequence of methylated data. It also predicts embryonic stem cells and baseline comparison to show genome downstream demonstration. Deep learning was mentioned to have outpace conventional techniques, as Kernes et al. (2016) used graph convolutions to encrypt molecular features, physical properties, and assay activities that permit potential collaboration of molecular encoded information.

**Deep learning used for medical imaging procedure**

Experts found DL relevant in diagnosing illnesses and interpreting medical images. The processes are conversantly enabled by CAD (computer-aided diagnosis) for assimilating the cause of diseases. CAD model helps to identify causes of neurological Alzheimers, sclerosis, and stroke progression through brain scans, multi-modal mapping of the infected region.

Over the years, convolutional neural network aids computer vision, especially the ability to personalize GPU to show parallel brain pathology (Havaei et al.2016)[29]. It further demonstrates CAD segmentation and shapes the analysis of the human brain. CAD has helped to overcome the challenges of difference in intensity & shapes of tumors and lesions using image protocols. Though, issues associated with CAD may include pathological tissue overlapping with healthy samples, RICIAN-Noise, non-Isotropic issues, and bias field effects evident in magnetic resonance images (MRI). Sometimes, the MRI cannot be handled automatically by CAD but requires a similar ML approach to decrypt data complexity and extract features through conventional approaches (Greenspan et al.2016)[30].

CNN as a deep learning approach works better than CAD in terms of data manipulation, operating patch images of abnormal tissues (e.g. CNNs medical imaging for lung diseases coordinated with computed tomography image). Experts used CNN and CT imaging applications to classify the manifestation of tuberculosis, the cell of the neural progenitor from somatic source, and hemorrhage color Fundus image detection. Yan et al (2016)[31] classified different anatomies with CT to understand human organ recognition using multistaged frameworks to extract patches of pre-trained stages. Jamaluddin et al.2016 seem it essential to use CNN for the segmentation of Isotense brain tissue and brain extraction through a multi-modality image tool. Avendi et al. 2016[32] described how DL algorithms encode deformable model parameters and facilitate left ventricle segmentation necessary for short-axis cardiac imaging. CNN tools have 2D image components for segmenting MRI & CT in 3D format, to eradicate issues found in anisotropic voxel sizing. CNN was adopted in orthogonal patch extraction to segment axis, sagittal, and corona view which reduces time and overfitting problems. (Fritscher et al.2016)[33].

Common limitations of CNN include its non-spatial dependencies and the need for pre-processing to bring conditional random fields. These issues can be altered by substituting with conventional ML approaches to solve problems of incomplete data training, limited annotated data, cost/time, and manual medical image annotations.

Previously, manual annotation was accepted to help the detection of medical images, but crowdsourcing according to (Greenspan et al.2016)[30] is a viable alternative due to its affordability, error-free medical image analysis. Havaei et al. 2016[29] used a transfer learning and fine-tuning approach to alleviate incomplete training issues on CNN, allowing the pre-trained data to be labeled manually. Tajbakhsh et al. (2016)[53] described the similarity between natural images and medical images by using the fine-tuning process to repeat the same experiment. Shin et al. (2016)[34] applied transfer learning to natural images of a thorax-lymph node to detect lung disease and the result shows consistency in performance without losses.

Chen et al (2015)[35] identified fetal abdominal standards with a transfer learning approach to display low-layer CNN pre-training effects on natural images while implementing multi-tasking to handle the CAD image imbalance. Cheng et al.2016[36] utilized denoising autoencoder to diagnose breast legions and pulmonary nodules in CT scans. Shan et al. (2016) [37]tried Stack Sparse Autoencoder on Microaneurysms Fundus images to detect diabetic retinopathy whilst uses Softmax Output Layer to show Alzheimer's disease prediction with functional magnetic resonance images (fMRI). Li et al. (2015)[38] used the RBM method to effect biomarkers from MRI and position emission tomography (PET SCAN) and the result of the scan shows 6% accuracy. Kuang et al. (2014)[39] discriminate attention deficit hyperactivity disorder with the same FMRI application.

Hence, experts extracted RBM latent hierarchical 3D patch features from the brain using image segmented tools and Brosch et al. (2013)[40] implored manifold learning method to study 3D brain images, its full automated shape, and cranial nerves. Deep learning methods are known to outpace conventional approach through low-contact optic tracts and other human pathological anatomies. Beaulieu-Jones et al. 2018 [60] found pipeline models relevant in detecting & segmenting objects to achieve an automatic volumetric image process called marginal space DL. MSDL handled hierarchical marginal spacing with automatic features to detect deep learning datasets.

**Literature review**

Unsupervised learning techniques are characterized by an unlabeled dataset using metrics of low-high dimensional subspace anomalies for detecting clusters of data. (E.g. Prediction of heart & hepatitis diseases). Collins and Yao, 2018 [43] defined prognosis as the method of predicting disease with clinical practice settings whilst showing multi-modal data. Wang et al. 2012 [44] use prognosis to diagnose diabetics registered in electronic health data records. Medical image analysis follows enhancement, detection, classification, and segmentation procedure to reconstruct and store data. Chen et al. 2017 [45] implemented the reconstruction of MRI and CT image datasets using generative adversarial networks (GANS).

The generative adversarial network (GAN) offers MRI reconstruction by cleaning motion pictures, artifacts, and handling image fusion & registration. El-Gamal et al. (2016)[46] noted the significance of image registration in surgical spine implant, tumor removal, and neuro-surgical process. 56 developed an image registration framework called "Quick-Silver", for large deformation mapping and diffeomorphic metric prediction. Before image registration, data retrieval enables physicians to check the large images of patients' repeated visits to the clinic. Zech et al. 2018 [47] shared natural language processing methods for annotating retrieved images from clinical radiological reports.

To achieve real-time health monitoring with DL wearables, IOT sensors, and smart devices, DL clouds must be integrated into smart devices to attain the required results. DL had found its place in clinical workflows for predicting & diagnosing diabetes, dengue, heart & liver diseases, whereby IBM advanced CAD system to CADX that displays automatically fatty liver in Kurtosis image (MA et al. 2016)[48].

Zhang, 2019 [49] utilized clinical reinforcement learning to study the optimal diagnosis and treat patients by characterizing its performance evaluation with different methods (Q value iteration, tabular learning, Q-iteration, and deep Q-learning). The RL method helps to treat sepsis in intensive care units. The same clinical time-series data were studied to provide medical intervention to patients in intensive care units by using CNN and LSTM to predict traumatic brain damage, estimate the mean-variance of arterial blood pressure and intracranial pressure monitoring (Rau, 2018) [50].

Recently, experts have adopted Attention Models to forecast ICU activities, integrate multivariance time-series measurement, and controlling unexpected respiratory issues. To control NIP challenges, Neveol, 2018[52] used the CLAMP Toolkit to monitor different states of clinical text analysis of language acronyms, disparity, and quality variance. Several doctors studied clinical documentation, especially on how to use clinical speech and audio processing to minimize time spent on administrative tasks and medical reports. (Wallace, 2019)[51]. Speech and audio processing serve the purpose of identifying disorders using vocal hyper-functional tools to review patients with dementia and Alzheimers.

**Limitations**

Privacy and security challenges are primary limitations discovered in deep learning algorithms. The issue of data collection vulnerabilities is experienced during DL adoption of large datasets which also consumes time and human efforts. The problem of incorrect or altered datasets can lead to wrong diagnoses. Latif et al. 2018 [54] disclosed that instrumental and environmental noise from smart machines can cause an unnecessary disturbance, especially MRI multi-shots, high sensitive modal-motion, and an increased risk of mis-diagnosis due to mistakes in artifacts; can be detrimental to human health.

Unqualified physicians without knowledge of data analytics can commit unforgivable errors in medical diagnosis. Caruana et al. 2015 [55] explained the difficulties in data labeling and annotation while 86 lamented about ambiguous medical image classification which may lead to confusion and disagreement between physicians. Xia et al. 2012 [56] indicate that the use of inappropriate algorithms can be detrimental and life-threatening; if improper annotation happens while suggesting the use of meticulous approaches in the labeling of datasets to limit inefficiencies. Due to limited or imbalance datasets, wrong diagnoses can cause death of millions, and missing data sparsity values can lead to unmeasured or repetition in taking samples. Biggio et al. 2012 [57] viewed model training vulnerabilities, as improper training or incomplete breach of privacy causing model poisoning and data theft. To corrupt an already collected data is known as "data poisoning" which requires security, especially during digital forensics & bio-metrics. In case of compromise during deep learning deployment, realistic healthcare settings will experience distribution shifts, leading to incomplete data vulnerability in the testing phase.

**Security and recommendations**

Numerous security threats, influence & violations are associated with DL algorithms, constituting to integrity attack, and other vulnerabilities. Such an act can destroy the progress of DL in health and other fields of science. Usama et al. 2019[58] mentioned adversarial machine learning vulnerabilities inserted in input samples to evade privacy and data integrity.

A data breach can cause modal poisoning & privacy evasion; whereby clinical deep learning applications are constantly under an attack. However, safety, privacy, ethical regulations & policy are being reinstalled to ensure the quality of data exchange standards.

David et al. (2015)[59] recommends hyper-plane commodity data cryptography to control data breach in naïve Bayes classifiers. Zhu et al. (2018)[66] suggest the use of polynomial aggregation and random masking protect SVM with non-linear kernel algorithms.

Jagieiski et al. (2018) indicate that data privacy can be reassured with a TRIM tool to protect linear creations. Lui et al. Ascertained that DL frameworks can be secured with XMPP serve while Malalhi et al (2019[68] named Paillier homomorphic encryption for security Naïve Bayes, SVM neural network and FKnn-CBR used to rescue liver patients in India hospitals. Takabi et al. (2016)[69] suggest homomorphic encryption for deep neural networks that control more than 15 datasets in repositories. To guarantee the safety of logistic regression, Kim et al. (2018) [70] recommends homomorphic encryption to secure medical binary datasets. To update healthcare infrastructure, Finlayson et al. 2019 [71] suggest the use of the international classification of disease system which helps to minimize dataset vulnerabilities. However, privacy can be preserved with a cryptographic approach, homomorphic encryption, garbled circuiting, and secured processors. The Intel SGX processor offers confidentiality and authorized access to systems like K-mean, decision trees, and SVM (Ohrimenko et al. 2016)[72]. Google Inc added federated learning method to distribute data, decentralize scheme and predict heart-related diseases. McMahan et al. 2017 [73].

To control adversarial attacks, it's important to modify models using defensive distillation, network verification, gradient regularization, and classifier robustification.

**Conclusion**

Deep learning and health informatics algorithms will continually expand to other branches of science, as wearable smart monitoring devices are presently used to track and diagnose Parkinson's diseases. Google Glass conducted prototype child therapeutic analysis to monitor and diagnose autism spectrum disorder. To preserve mental health, psychiatric hospitals are screened, diagnosing, and monitoring depression with a system-on-chip solution that accelerates filters, and reveals heart rates on ECG. Smart monitoring devices are unique, compatible, embedded with DL, and simple to operate. However, aging adults may find it challenging.

The future of DL and health informatics depends on the recent 5G wireless network, proposed to bring about new devices for testing red protein (Hemoglobin); especially for transporting oxygen to the blood. The accuracy of clinical results can be validated with a cross-validation approach to unravel the results.

27. Da He, D., Winokur, E. S., & Sodini, C. G. (2012). An ear-worn continuous ballistocardiogram (BCG) sensor for cardiovascular monitoring. Paper presented at the Engineering in Medicine and Biology Society (EMBC), 2012 Annual International Conference of the IEEE.

28. Winokur, E. S., Delano, M. K., & Sodini, C. G. (2013). A wearable cardiac monitor for long-term data acquisition and analysis. IEEE Transactions on Biomedical Engineering, 60(1), 189-192.

29. Havaei, M., Guizard, N., Larochelle, H., Jodoin, P.: Deep learning trends for focal brain pathology segmentation in MRI," CoRR, vol.abs/1607.05258, (2016).

30. Greenspan, H.,Van Ginneken, B., Summers, R. M. :Guest editorial deep learning in medical imaging: Overview and future promise of an exciting new technique," IEEE Trans. Med. Imag., vol. 35, no. 5, pp.1153–1159, May (2016).

31. Yan, Z., Zhan, Y., Peng, Z., Liao, S., Shinagawa, Y., Zhang, S., Metaxas, D. N., Zhou, X. S. :Multi-instance deep learning: Discover discriminative local anatomies for body part recognition," IEEE Trans.Med. Image, vol. 35, no. 5, pp. 1332–1343, (2016).

32. Avendi, M., Kheradvar, A., Jafarkhani, H.: A combined deep-learning and deformable-model approach to fully automatic segmentation of the left ventricle in cardiac MRI," Medical image analysis, vol. 30, pp. 108–119, (2016).

33. Fritscher, K., Raudaschl, P., Zaffino, P., Spadea, M. F., Sharp, G. C., Schubert, R. :Deep neural networks for fast segmentation of 3dmedical images," in MICCAI, (2016), pp. 158–165.

34. Shin, H.C., Roth, H. R., Gao, M., Lu, L., Xu, I., Nogues, J., Yao, D., Mollura, R., Summers, M. :Deep convolutional neural networks for computer-aided detection: Cnn architectures, dataset characteristics, and transfer learning," IEEE Trans. Med. Imag., vol. 35, no. 5, pp.1285–1298, (2016).

35. Chen, H., Ni, D., Qin, J., Li, S., Wang, X. T., Heng, P. A.: Standard plane localization in fetal ultrasound via domain transferred deep neural networks," IEEE J. Biomed. Health Inform. vol. 19, no. 5, pp. 1627–1636, (2015).

36. Cheng, Z., Ni, D., Chou, Y.H., Qin, J. C., Tiu, M., Chang, Y.C., Huang, C. S., Shen, D., Chen, C. M.: Computer-aided diagnosis with deep learning architecture: Applications to breast lesions in our images and pulmonary nodules in ct scans," Scientific reports, vol. 6, (2016).

37. Shan, J., Li, L.: A deep learning method for micro-aneurysm detection in fundus images," in IEEE CHASE, (2016), pp. 357–358.